\documentclass[11pt]{article}    

\usepackage{verbatim}

\usepackage{fancybox}
\usepackage{amsmath}
\usepackage{amssymb}
\usepackage[dvips]{graphicx}
\usepackage{psfrag}
\usepackage{epic}
\usepackage{eepic}
\usepackage[square,sort,comma,numbers]{natbib}
\usepackage{cleveref}

\usepackage{algpseudocode,algorithm,algorithmicx}

\DeclareGraphicsExtensions{ps}
\graphicspath{{fig/}{../fig/}{tmp/}}
\makeatletter
  \newcommand\figcaption{\def\@captype{figure}\caption}
  \newcommand\tabcaption{\def\@captype{table}\caption}
\makeatother

\newcommand{\creflastconjunction}{, and~}

\def\mi{\begin{equation}}
\def\mf{\end{equation}}
\newcommand{\ud}{\mathrm{d}}

\renewcommand{\v}[1]{\ensuremath{\mathbf{#1}}} 
\newcommand{\gv}[1]{\ensuremath{\mbox{\boldmath$ #1 $}}}  

\newcommand{\transp}{\top}

\def\mean{\mathcal{E}}

\RequirePackage{color}
\newcommand{\linelabel}[1]{}
\newcommand{\ma}[1]{#1}

\title{Model uncertainty estimation using the expectation maximization algorithm and  a particle flow filter}

\author{Mar\'{i}a Magdalena Lucini$^{a,b}$ and Peter Jan van Leeuwen $^{a,c}$
  and Manuel Pulido$^{a,b}$\\
\scriptsize$^a$ Data Assimilation Research Centre, Department of Meteorology, University of Reading, UK\\
\scriptsize$^b$ FaCENA, Universidad Nacional del Nordeste and CONICET, Argentina\\
\scriptsize$^c$ Department of Atmospheric Science, Colorado State University, USA}

\begin{document}

\maketitle

\begin{abstract}
  Model error covariances play a central role in the performance of data
  assimilation methods applied to  nonlinear state-space models. However,
  these covariances are largely unknown in most of the applications. A misspecification of the model
  error covariance has a strong impact on the computation
  of the posterior probability density function, leading to unreliable
  estimations and even to a  total failure of the assimilation procedure.  In
  this work, we  propose the combination of the Expectation-Maximization
  algorithm (EM) with  an efficient particle filter to estimate the model
  error covariance, using a batch  of observations.  Based on the EM algorithm
  principles, the proposed  method  encompasses two stages: the  expectation
  stage, in which a particle filter  is used with the present  estimate of the
  model error covariance as given to  find the probability  density function
  that maximizes  the likelihood,  followed by a maximization stage in which
  the expectation under the probability  density function found in the
  expectation step is maximized as  a function of the elements of the model
  error  covariance.
 This novel algorithm here presented combines the EM with a fixed
 point algorithm and does not require a particle smoother to approximate the
 posterior densities.  We demonstrate that the new method accurately and efficiently solves 
the linear model problem. Furthermore, for the chaotic nonlinear
Lorenz-96 model the method is stable even for observation error covariance 10 times larger
than the estimated model error covariance matrix, and also that it is successful in high-dimensional 
situations where the dimension of the estimated matrix is 1600.
\end{abstract}

{\bf Keywords:}
  Particle Filters, State-Space Models, Model error covariance, EM algorithm


\section{Introduction}

Several research areas in which the object of research is a complex system,
such as for instance the atmosphere, the ocean and biological systems, require to estimate the state of the system through partial
observational information which is distributed in time. Surrogate models of
the system, which represent approximately the time evolution of the variables,
are included as a source of information in the state estimation. This has a twofold
aim, it regularizes the state estimation at a time and propagates information
of the system and its uncertainty to the subsequent times. This estimation
problem was called data assimilation in geophysical sciences
\cite{book:daley1993, book:law2015data}, and this terminology has then been popularized in other areas.  The
so-called 
state-space model is composed by two stochastic equations, one represents the evolution of the hidden state
variables, henceforth referred as dynamical model,  and the other one maps the state of the system into the
observations space through an observational model. 

The increase of observational data availability, particularly indirect
observations, and the increased complexity in surrogate models, introduce
nonlinear dependencies in the dynamical and observational models which in turn
lead to non-Gaussian statistics. To account for these non-Gaussian statistics in
high-dimensional state-space models is essential for  the inference and
represents one of the major challenges in the area
\cite{art:Dowd_etal_14,art:Ueno_etal_10, art:Wikle_etal_07}. Gaussianity-based techniques such as all the variants of the Kalman
filter and optimization techniques based on maximum a posteriori estimation,
also known as variational data assimilation, cannot deal with strongly non-Gaussian
statistics. Monte Carlo techniques  are one of the most promising methodologies
that can fully consider the  non-Gaussian uncertainty in the sequential inference
\cite{doucet2009tutorial,proc:Gordon_etal_93, book:Robert_Casella}. These techniques aim to represent
the state distribution by a set of realizations (referred as particles) of this
distribution. Particles are evolved between estimation times by the use of the dynamical model. However, one contention point in these
techniques is the particle degeneracy along the time sequence. After a few cycles, most
of the particles finish with negligible weights, i.e., their  likelihood to the
observation is low or null, and only one particle remains with weight 1. This limitation
is particularly important in high-dimensional systems. One
solution is resampling, however it brings another limitation since only a few particles with large weight remain
and therefore the diversity is lost, particularly in experiments which can only afford a small number of particles.

Recently a new framework for particle filters, called particle flow filters,
has been proposed \cite{art:Bunch_Godsill_16,proc:Daum_etal_10}. Within this framework, 
particles are moved from the initial proposed distribution to the posterior
distribution using flows that are consistent with Bayes rule. Particle flows potentially avoid the need of resampling since the
particles are moved to regions of higher likelihood  of the state space. The
flow is not unique and requires the solution of an ordinary differential
equation under a given regularization. Alternatively, if the Gaussian
assumption is taken,  the expression for the flow can be derived analytically
and the flow is unique and known as  Gaussian particle flow. In a recent work,
Pulido and vanLeeuwen \cite{art:Pulido_MPF_19} showed that under the
assumption that the flow lies in a reproducing kernel Hilbert space, then it
is uniquely determined through a Monte Carlo integration, an interacting
particle system. This approach combines Monte Carlo sampling with variational inference. It is shown that this sequential Monte Carlo filter does not require resampling even for long time sequences and can work  in relatively high-dimensional systems.

A crucial assumption in particle filters is that the model error is known. The
model error uncertainty is included in the evolution of the particles and is
essential to account for the prediction uncertainty.  This is particularly the
case in particle flow filters, in which the sequential prior density is
assumed to be known. In practice, model error
is not known and it is highly dependent on the surrogate model we are
using. The structure of the model error, in terms for example of correlation between variables, is expected to be different if different surrogate models are used. Therefore, there is a current need for the development of model error estimation techniques that may be applied for sequential Monte Carlo filters.

Likelihood-based methods for model error covariance estimation may be broadly classified into two
groups: maximum likelihood estimation and Bayesian inference. Maximum
likelihood estimation methods have been applied 
assuming   that covariance parameters are deterministic
\cite{book:Cappe_etal_06, art:Kantas_etal_15, art:Ueno_etal_10, art:Ueno_Nakamura_14}. In this
case, non a priori information on the parameters is required. When the
gradient of the likelihood can be obtained analytically, Newton-Raphson
optimization methods may be applied to maximize the likelihood function. When
this is not feasible one
of the most popular methods for maximum likelihood estimation is expectation
maximization (EM) \cite{art:Dempster_etal_77}. One of the reasons of its
widespread use is that it is readily applicable. In particular, contrary to
Newton-Raphson optimization methods, expectation-maximization does not depend
on (optimization) parameters. The second group is a Bayesian approach in which
the model error covariance parameters are interpreted as stochastic and the
prior distribution of the parameters needs to be given. In this case, some
hypotheses on the correlations between state space variables and parameters
are required in high-dimensional systems \cite{art:Santiti_Jones_15}. Some authors \cite{art:Matisko_Havlena_13,art:Stroud_etal_07} assume that the uncertainty in the parameters does not affect the state density, while \cite{art:Scheffler_etal_18} use the marginalization of the hidden state for the parameter estimation in a hierarchical Bayesian framework.

In this article, we propose a new method based on a maximum likelihood approach
to estimate the model error
covariance matrix in state-space models. The article is organized as follows.  In 
\cref{sec:methodology} we formulate the problem  giving a brief
overview of particle filters and some details of a recently proposed particle
flow filter to be used in the experiments before describing the proposed method to estimate the
model error covariance matrix in \cref{sec:inference}. Further details about
its derivation are
given in  \cref{app} and \cref{sec:derivQ}. The proposed method is tested on a
simple autoregressive linear model and on the Lorenz-96 model with 8 and 40
dimensions. Experiments design, results, and comparison with existing methods
are shown in \cref{sec:experiments} while in  \cref{sec:conclusions} we present
some conclusions.

\section{Methodology}\label{sec:methodology}
\subsection{Problem formulation}\label{sec:problem-formulation}

Consider a state-space model, consisting of a non-linear system  and a non-linear observation model, described by

\begin{align}\label{eq:model}
        {\bf{x}}_k &= {\mathcal{M}}({\bf x}_{k-1}) + \gv\beta_k\\
        {\bf{y}}_k &= {\mathcal{H}}({\bf{x}}_{k}) + \gv\epsilon_k
\end{align}
where ${\bf{x}}_k \in \mathbb{R}^{N_x}$, called the state vector, is a hidden or
latent process, $\{{\bf y}_k\}_{k=1}^K$ is a time series of observations measured
at times $k=1,\ldots,K$ with ${\bf y}_k \in \mathbb{R}^M$.  The maps $\mathcal{M}(.)$ and $\mathcal{H}(.)$ denote the
non-linear deterministic dynamical model and the (possibly non linear) observation
operator, respectively. The state process is assumed to be a Markov process, whereas the conditional density of the observations  ${\bf y}_k$  depends only on the current state $\v x_k$ for $k = 1, \ldots,K$.
The stochastic term  $\gv\beta_k$ accounts for 
the missing physics in the model and its numerical approximations, whereas $\gv\epsilon_k$ is the observation 
noise. In this work, we assume that observational and model uncertainty are
additive and $ \gv\beta_k \sim \mathcal{N}(0, {\bf Q}_k)$
and  $\gv \epsilon_k \sim \mathcal{N}(0,\v R_k)$, where $\v Q_k$ and $\v R_k$
belong to the subspace of positive definite matrices in $\mathbb{R}^{N_x\times N_x}$ and
$\mathbb{R}^{M\times M}$, respectively,  representing 
the model error  and observation error covariance matrices at
time $k$. We denote by $\gv \theta_k \in \gv \Theta$ the vector of parameters
$(\v Q_k, \v R_k)$.

Since $\mathcal M$ and $\mathcal H$ are surrogate models which approximate the
system evolution and the processes that relate the observations with the hidden state, the model error and observational error covariances are expected to be
largely  unsconstrained physically. Observational errors may be partially
constrained from the knowledge of measurement errors, whereas representation
errors, arising from the fact that model and observations often represent
reality differently, are hard to determine in practice. Therefore,  the
parameters $ \gv \theta_k$ of the state space model are unknown and need to be
inferred from the data as well as the hidden state $\v x_k$. In principle,
unknown parameters from the dynamical and observational models may also be included
in $ \gv \theta_k$ \cite{art:Pulido_etal_18}; in this work we consider these parameters are provided.

We  assume that the model error covariance $\v Q_k$ and the observation error covariance $\v R_k$  vary slowly
within $K$-cycles $\,$ (the temporal scale of covariance variations is longer that $K$-cycles), that is $\v Q_k = \v Q$, $\v R_k = \v R$,
$\forall k=1,\ldots, K$  and  propose a method based on a time-batch of observations to estimate the
model error covariance $\v Q$ for particle filters (PF).
The information provided  by the observations
along the $K$ times, $\v y_{1:K}$, where the subindices $1:K$ denote the set
$\{{\bf  y}_1, \ldots, {\bf y}_K\}$, is considered essential to
regularize/constrain the $N_x \times N_x$ unknowns from  $\v Q$. The coupling  between observations at different times is
  produced through the dynamical model $\mathcal{M}$.

The method here presented is based on maximum likelihood estimation: given a
set of independent  observations
$\{{\bf{y}}_k, k=1,\ldots,K\}$  from a probability density function represented
by $p({\bf y}_{1:K};\gv  \theta)$, a nonlinear
dynamical model $\mathcal{M}$ and an observation operator $\mathcal{H}$, we
seek to maximize the likelihood of the observations as a function of the
statistical parameters $\v Q$ given an observation error covariance $\v R$  in the presence of a hidden state $\v x_{0:K}$.

\subsection{Particle filters}

State-space models are generally used in sequential data assimilation to
estimate or reconstruct the hidden state ${\bf x}_k$ given the
observations 
${\bf y}_{1:K}$. This can be done by computing the filter densities
$\{p(\v x_k|\v y_{1:k})\}_{k=1:K}$ or smoother densities
$\{p(\v x_k|\v y_{1:s})\}_{k=1:K}$ with $s \geq k$. Having  prior
knowledge of the initial state ${\bf x}_0$, that is, given a prior
background probability density function (pdf) $p({\bf x}_0)$, the
posterior pdf of a filter is the probability of the model state at time $k$, given all the available
information up to time $k$. In a Markovian system with observations ${\bf y}_k$ that are conditionally independent given the state, the filter
densities can be computed  recursively  using Bayes' rule to obtain the
posterior pdf

\begin{equation}\label{eq:filtering}
  p(\v x_k |\v y_{1:k};\gv \theta) = \frac{p({\bf y}_k|{\bf x}_k;\gv \theta)p({\bf x}_k|{\bf y}_{1:k-1};\gv \theta)}{p({\bf y}_k|{\bf y}_{1:k-1};\gv \theta)},
\end{equation}
where the prior pdf  $p({\bf x}_k|{\bf y}_{1:k-1};\gv \theta )$ is the forecast or
prediction pdf, $p({\bf y}_k|{\bf x}_k;\gv \theta)$ is the observation likelihood  defined by the observation model $\mathcal{H}$  and the
distribution of the  observation error $\gv \epsilon_k$, and  $p({\bf y}_k|{\bf y}_{1:k-1};\gv \theta)$ is a normalizing factor. Note that we consider here the
marginalized posterior pdf, in which the only state variable that is estimated
is the current one, considering all the past and the current observations. This assumption is essential when dealing with high-dimensional systems.

When the dynamical and observational models are linear and their
errors Gaussian, the filter densities are Gaussian and  the state can be computed using Kalman recursive
algorithms. However, for nonlinear models and/or nonlinear observational functions it is not 
possible to get a known distribution or a closed form for these filter
pdf's, and they should be somehow approximated. Classical particle filters
\cite{art:Arulampalam_etal_02, doucet2009tutorial,art:vanLeuween_09} are based on sequential importance sampling and resampling algorithms and  provide different methods to approximate these pdfs.

The basic idea behind a particle filter is to represent the posterior pdf $p(\v x_k|\v y_{1:k})$ by a
set of $N_p$ particles $\{x^{(j)}_k\}_{j=1:N_p}$ with
corresponding weights $\{w_k^{(j)}\}_{j=1:N_p} $ such that
$\sum_{j=1}^{N_p}w_k^{(j)}=1$.
That is, at time $k$, $p(\v x_k|\v y_{1:k};\gv \theta)$ is approximated by

\begin{equation}\label{eq:filPF}
 p(\v x_k|\v y_{1:k};\gv \theta) \doteq \sum_{j=1}^{N_p} w_k^{(j)}\delta({\bf x}_k - {\bf x}_k^{(j)})
\end{equation}
where $\doteq$ denotes approximation by an ensemble of particles and
$\delta(\cdot)$ is the Dirac $\delta$ function. 

Initially, a  set of $N_p$ particles$\{x^{(j)}_0\}_{j=1:N_p}$ with
corresponding weights $\{w_0^{(j)}=\frac{1}{N_p}\}_{j=1:N_p} $ is drawn from
the  prior pdf $p({\bf x}_0)$. These particles are sequentially evolved
in time using  a forecasting, weighting and resampling
scheme to obtain $\{x^{(j)}_k\}_{j=1:N_p}$ and
$\{w_k^{(j)}\}_{j=1:N_p}$ at each time step $k$. Different PFs were proposed 
depending on the  resampling,
forecasting or weighting approaches taken \cite{art:Arulampalam_etal_02,art:Pulido_MPF_19, art:Zhu_etal_16}.

\subsubsection{Variational mapping particle filter} \label{sec:MPF}

The variational mapping particle filter,VMPF, is a particle filter which is based
on optimization and Monte Carlo sampling. The particles are moved
deterministically via a sequence of maps, based on the optimal transport
principle. The maps seek to minimize the Kullback-Leibler divergence (KLD) between the target density, i.e. the posterior density $p(\v x_k|\v y_{1:k})$, and an intermediate density $q(\v x_k)$, that is the density represented through the sample, the set of particles, at a given cycle $k$. At the $i$-th optimization iteration, the KLD is given by
\mi
\mathcal D_{KL} ( q_{k,i}(\v x_k)\rVert p(\v x_k|\v y_{1:k}) ) = \int q_{i} (\v x_k) \log  \frac{q_{k,i} (\v x_k)}{p(\v x_k|\v y_{1:k})} \ud \v x_k, 
\mf
where the $q_{i}(\v x_k)$ is represented via $N_p$ sample points, i.e. particles, $\v x^{(1:N_p)}_{k,i}\sim q_{i}(\v x_k)$.

The density $q_{i}$ is the result of a map, $T_i(\v x_{k,i-1}) = \v x_{k,i-1}+\epsilon \v v_i(\v x_{k,i-1})$, which is a small perturbation to the identity
map ($\epsilon$ is assumed to be small). This means $q_{i}=T_i^\sharp q_{i-1}= q_{i-1}(T_i^{-1}) |\det J(T_i^{-1})|$. The maps are assumed to be in a reproducing kernel Hilbert space.

The optimal map that gives the steepest descent direction is shown  to be given by
\mi
\v v_i(\v x_{k,i-1}) = - \nabla \mathcal D_{KL}(\v x_k) =  \mean_{x'_k\sim q}\left[ \v K(\v x'_k, \v x_k ) \nabla_{x'_k} \log p(\v x'_k|y_{1:k}) +  \nabla_{x'}\v K(\v x'_k, \v x_k)\right], \label{KLgrad1}
\mf
where $\v K$ is a kernel (assumed here to be Gaussian) \cite{art:Pulido_MPF_19}.

Then  the optimization is a sequence of (sufficiently smooth) mappings for each particle $(j)$ along the steepest descent direction
\mi
\v x_{k,i}^{(j)}= \v x_{k,i-1}^{(j)} - \epsilon \nabla \mathcal D_{KL}(\v x_{k,i-1}^{(j)}).\label{mapping2}
\mf

If observational and model errors are assumed to be Gaussian, the gradient of the log-posterior density   is
\mi
\nabla_{\v x} \log p(\v x_k)=\v H^\transp \v R_k^{-1} \left(\v y_k- \mathcal{H}(\v x_k)\right)- \v Q_k^{-1} \left[\v x_k -\sum_{j=1}^{N_p}  \gv\beta_{k-1}^{(j)} \mathcal{M}(\v x^{(j)}_{k-1})\right], \label{gradp}
\mf
where $\gv\beta_{k-1}^{(j)} \triangleq \frac{w_{k-1}^{(j)} \psi^{(j)}_{\v Q_k}}{\sum_{m=1}^{N_p} w_{k-1}^{(m)}  \psi^{(m)}_{\v Q_k}}$, and $\psi^{(j)}_{\v Q_k}=\exp(-\rVert \v x_k - \v x_k^{(j)}\lVert^2_{\v Q_k})$. They could be
interpreted as weights of the forecast states which consider the distance
between the particles to the point under consideration. A more detailed
  description of the VMPF is found in \cite{art:Pulido_MPF_19}. Recently, a generalization to embed
  also the observational operator in the RKHS was proposed in \cite{art:Pulido_etal_19}

  One of the main advantages of the VMPF is that not only it efficiently samples
high- dimensional state spaces with a limited number of particles but also it
 does not suffer from sample impoverishment. 

\subsection{Parameter estimation}\label{sec:inference}

 Let us assume that $p(\cdot;\gv \theta)$ is a parametric distribution, with
 $\gv \theta \in \gv \Theta$, the parameter space. Given a set of 
 observations ${\bf y}_{1:K}=\{{\bf{y}}_k, k=1,\ldots,K\}$ taken along a time interval of length
$K$, a maximum likelihood estimation method aims at finding the value of
 $\gv \theta$ that maximizes the (incomplete) likelihood of the observations,

\begin{equation}\label{eq:incomplete-likelihood}
  L(\gv \theta) = p({\bf y}_{1:K};\gv \theta) = \int  p(\v x_{0:K},\v y_{1:K};\gv \theta)\mathrm{d}\v x_{0:K},
\end{equation}  
or equivalently, the log-likelihood function
\begin{equation}
l(\gv \theta)=\ln p({\bf y}_{1:K};\gv \theta)= \ln \left(\int  p(\v x_{0:K},\v y_{1:K};\gv \theta) \mathrm{d}\v x_{0:K} \right). \label{eq:observationlogLikelihood}
\end{equation}

The likelihood function \cref{eq:incomplete-likelihood} can be interpreted as
how probable the set of observations $\v y_{1:K}$ would be for different choices of $\gv \theta$.
An analytic form for the log-likelihood function is not achievable in practice, and
the numerical evaluation of \cref{eq:observationlogLikelihood} may involve
high-dimensional integrations, what is intractable. In some situations  the optimization task can
be accomplished by using numerical optimization routines like 
Newton-Raphson techniques to solve the nonlinear equations obtained by
differentiating the log-likelihood function
\cref{eq:observationlogLikelihood} \cite{ book:Cappe_etal_06,art:Gupta_Mehra_74,art:Pulido_etal_18}. However, even in these particular situations, other methods are
preferable due to the difficulty of implementing optimization methods and
tuning their parameters. Gradient optimization methods may be not stable
numerically for certain sets of parameters. 

The expectation-maximization  (EM)  algorithm \cite{art:Dempster_etal_77}, is a widely used numerical
method that aims at  maximizing the log-likelihood of the observations as a
function of the statistical parameters $\gv \theta$ in the presence of a
hidden state $\v x_{0:K}$  in successive
iterations without the need to evaluate the complete log-likelihood function.

It basically consists in maximizing iteratively an
intermediate function defined as
\begin{align}
  \mathcal{G}(\gv \theta', \gv \theta)& \triangleq \mathbb{E}_{\gv \theta'}\left[\ln p(\v x_{0:K},\v y_{1:K}; \gv \theta)\right]\label{eq:intermediate}\\
  & =  \int \ln p(\v x_{0:K},\v y_{1:K}; \gv \theta)p(\v x_{0:K}|\v y_{1:K}; \gv \theta')\mathrm{d}\v x_{0:K} \label{eq:intermediate2}
\end{align}
where $\gv \theta', \gv \theta \in \gv \Theta$. This intermediate function
$\mathcal{G}$ is, generally, much simpler to maximize than the incomplete
  log-likelihood defined in \cref{eq:observationlogLikelihood}.

  Starting from an initial parameter $\gv \theta_0$, the two steps of the EM
  algorithm at iteration $s$ can be summarized as:

  \begin{itemize}
     \item {\bf Expectation Step} (E-Step): Calculate the required densities
       to compute the intermediate function
       $\mathcal{G}(\gv \theta_{s-1},\gv \theta)$ as in  \cref{eq:intermediate2}.
     \item {\bf Maximization Step} (M-Step): find $\gv \theta_s = \displaystyle\max_{\gv \theta\in \gv \Theta}\mathcal{G}(\gv \theta_{s-1},\gv \theta)$.
  \end{itemize}

The assumptions of a hidden  Markov model and mutually independent
observations in \cref{eq:model} allow us to express the joint probability
density function $p(\v x_{0:K},\v y_{1:K}; \gv \theta)$ as

\begin{equation}\label{eq:joint_pdf}
 p(\v x_{0:K},\v y_{1:K}; \gv \theta) = p(\v x_0; \gv \theta)\prod_{k=1}^Kp(\v x_k|\v x_{k-1}; \gv \theta)\prod_{k=1}^Kp(\v y_k|\v x_k; \gv \theta).
\end{equation}

To estimate the parameters of the state-space model using the EM algorithm,
the expectation of this last pdf under the conditional (smoother) pdf
$p(\v x_{0:K}|\v y_{1:K}; \gv \theta')$ must be computed in the E-Step of the
algorithm. In the case of a
linear Gaussian model this can be accomplished by means of a Kalman smoother
\cite{art:Shum_Sto_82}.  This was further extended to the ensemble Kalman filter in \cite{art:Dreano_etal_17,art:Pulido_etal_18}.

When the dynamical or observational model are non-linear and therefore  the
joint density non-Gaussian, the expectation in equation \cref{eq:intermediate} may be
intractable and a  different approach must be taken. A generally used approach
is to approximate the expression in \cref{eq:intermediate} by  generating
samples of the smoother pdf $p(\v x_{0:K}|\v y_{1:K};\gv \theta')$ using a
particle smoother \cite{art:Briers_etal_10, art:Kantas_etal_15, art:Olsson_etal_08}. However, the use of  particle smoothers in data
assimilation represents a  computational challenge, since they not
only tend to degenerate rapidly but also have a poor performance in moderate
to high dimensional spaces, particularly if the time sequence is long (large $K$).

The requirement of a particle smoother in the E-Step of the EM algorithm is
due to the fact that the likelihood of the observations $p(\v y_{1:K};\gv \theta)$
is usually obtained by marginalizing the joint pdf $p(\v x_{0:K},\v
y_{1:K};\gv \theta)$ over the whole state $\v x_{0:K}$ (cf. equation \cref{eq:incomplete-likelihood}). 


Instead of using this last expression for the likelihood of the observations,
and following the notation of \cite{art:Carrassi_etal_17}, the
likelihood of the observations (model evidence) can be decomposed as $p(\v y_{1:K};\gv \theta) =\prod^K_{k=1}p(\v y_{k}|\v y_{k-1};\gv \theta)$, with the convention $\v y_{1:0}=\{\emptyset \} $. Marginalizing this last
expression we obtain

\begin{align} 
  p (\v y_{1:K};\gv \theta) &=\prod^K_{k=1} p(\v y_{k}|\v y_{k-1};\gv \theta) \nonumber \\
  &= \prod^K_{k=1} \int   p(\v y_{k}|\v x_{k};\gv \theta) p(\v x_{k}|\v y_{1:k-1};\gv \theta) \mathrm{d}\v x_k \label{eq:likelihood2}
\end{align}
and therefore we can rewrite the logarithm of the incomplete likelihood  \cref{eq:incomplete-likelihood} as
\begin{equation} \label{eq:loglikelihood2}
l(\gv \theta)=\log p (\v y_{1:K};\gv \theta) = \log \prod^K_{k=1} \int p(\v y_{k}|\v x_{k};\gv \theta) p(\v x_{k}|\v y_{1:k-1};\gv \theta) \mathrm{d} \v x_k.
\end{equation}

Using this last expression  instead
of the commonly used expression  \cref{eq:observationlogLikelihood},  the
intermediate function $\mathcal{G}(\gv \theta',\gv \theta)$ of the EM algorithm
can be written as

\begin{align}
\mathcal{G}(\gv \theta',\gv \theta )&= \sum_{k=1}^K\int p(\v x_{k}|\v y_{1:k};\gv \theta')\log\left(\frac{p(\v y_{k}|\v x_{k};\gv \theta) p(\v  x_{k}|\v y_{1:k-1};\gv \theta)}{p(\v x_{k}|\v y_{1:k};\gv \theta')}\right) \mathrm{d} \v x_k \label{itermFn1}
\end{align}
as it is shown in \cref{app}. Note that this last expression for
$\mathcal{G}(\gv \theta',\gv \theta )$ is written in terms of filter and
forecast pdf's,  while smoother pdf's are no longer required. As the denominator of \cref{itermFn1} does not depend on $ \gv \theta$,  maximizing
$\mathcal{G}(\gv \theta', \gv \theta)$ w.r.t $ \gv \theta$ is equivalent to maximizing

\begin{align}
  \label{eq:int-theta}
\mathcal{G}( \gv \theta', \gv \theta)&=  \sum_{k=1}^K\int p(\v x_{k}|\v y_{1:k};\gv \theta')
\log\left(p(\v y_{k}|\v x_{k};\gv \theta) p(\v x_{k}|\v
  y_{1:k-1};\gv \theta)\right) \mathrm{d} \v  x_k  \\
  &=  \sum_{k=1}^K\mathcal{E}_{\gv \theta'}\left[\log(p(\v y_{k}|\v x_{k};\gv \theta) p(\v x_{k}|\v  y_{1:k-1};\gv \theta))\right].
\end{align}

So far $\gv \theta$ denotes the set comprised by $(\v Q, \v R)$, the model error and
observation error covariance matrices, respectively. As there
is a certain degree of knowledge about the instruments noise and how
observations are measured or obtained,
$\v R$ is usually determined empirically in practice by estimating these noises and the
errors between the state and observation space. However, this is far from
being an accurate representation of $\v R$ and  during the last few years
a great effort has been dedicated to the study and  estimation of the
observation error covariance matrix, and many works have been
published on these topics
(\cite{art:Stewart_etal_13,art:Ueno_Nakamura_14,art:Ueno_Nakamura_16,art:Waller_etal_17} and references
therein). Some of these works focus on studying the most plausible structure for
$\v R$ given the nature of how observations were measured taking into account
correlations between observation errors whilst some others just propose a fixed
structure for $\v R$ and a methodology to estimate it. What these different approaches have in common is that
they assume that the model error covariance matrix is already given, or
known. The model error covariance matrix $\v Q$ is,
perhaps, the most difficult one to estimate or determine, since it accounts
for the model inaccuracies and deficiencies in representing the missing underlying
physics, the errors in parameterizations, the unresolved and smaller scales
and the numerical schemes used. 
Some works have been devoted to the joint estimation of
$(\v Q, \v R)$ and an up to date and detailed review of these techniques is presented in
\cite{art:review_tandeo}.

The main purpose of this work is to provide a method to estimate $\v Q$, the covariance matrix of
the model error $\gv\beta$. Assuming that $\v R$ is known, replacing $\gv \theta =  \v Q $ and having in mind
that by hypothesis the density $p(\v y_{k}|\v x_{k};\v Q)$ is assumed to be
independent of $\v Q$, then starting from an initial
guess $\v Q_0$, the two steps of the
EM algorithm at iteration $s$ can be summarized  as:

  \begin{itemize}
     \item {\bf Expectation Step} (E-Step): Calculate the required densities
       to compute the intermediate function
       \begin{align} \label{eq:interQ}
         \mathcal{G}(\v Q_{s-1},\v Q)=  \sum_{k=1}^K\mathcal{E}_{\v Q_{s-1}}\left[\log( p(\v x_{k}|\v y_{1:k-1};\v Q))\right].
        \end{align} 
     \item {\bf Maximization Step} (M-Step): find $\v Q_s=\displaystyle\max_{\v Q \in \gv \Theta}\mathcal{G}(\v Q_{s-1},\v Q)$,
  where $\gv \Theta$ is the space of positive definite matrices of order $N_x$.
  \end{itemize}
Using a particle filter, the posterior pdf $p(\v x_{k}|\v y_{1:k};\v Q_{s-1})$ can be
approximated  by a set of particles and their corresponding weights as in
equation  \cref{eq:filPF} with $\gv \theta =\v Q_{s-1}$.

As  we assume a Gaussian
model noise $\gv\beta \sim \mathcal{N}(0, {\bf Q})$
(see  \cref{sec:problem-formulation}), then the transition density is
given by $p(\v x_k|\v x_{k-1}; \v Q) = \phi(x_k , \mathcal{M}(\v x_{k-1}), \v Q)$
where

\noindent $\phi(x_k , \mathcal{M}(\v x_{k-1}), \v Q) \triangleq \frac{1}{(2\pi)^{n/2}|\v Q|^{1/2}}\exp\left\{-\frac{1}{2}(\v x_k -\mathcal{M}(\v x_{k-1}))^\top\v Q^{-1}(\v x_k-\mathcal{M}(\v x_{k-1}))\right\}$.

Therefore, the  prediction density $p(\v  x_{k}|\v y_{1:k-1};\v Q)$ can be approximated by

\begin{align} \label{eq:forecast}
  p(\v  x_{k}|\v  y_{1:k-1};\v Q) \doteq  \sum_{i=1}^{N_p} w_{k-1}^{(i)}\gv \phi\left(\v x_k,\mathcal{M}(\v x_{k-1}^{(i)}),\v Q \right),
\end{align}
where $w_{k-1}^{(i)},\v x_{k-1}^{(i)} $ are the $i$-th weight and particle,
respectively, obtained by a particle filter at time step $k-1$.

Combining these approximated pdf's and computing the required expectations, 
the intermediate function
$\mathcal{G}(\v Q_{s-1},\v Q)$ given in equation \cref{eq:interQ} can now be written in
terms of $N_p$ particles as,

\begin{align} \label{eq:interQ-PF}
         \mathcal{G}(\v Q_{s-1},\v Q) \doteq  \sum_{k=1}^K
         \sum_{j=1}^{N_p}w_{k, \v Q_{s-1}}^{(j)} \log \left(\sum_{i=1}^{N_p}
         w_{k-1, \v Q}^{(i)}\ \ \gv \phi\left(x_k^{(j)},\mathcal{M}(\v x_{k-1}^{(i)}),\v Q\right) \right).
        \end{align} 

Differentiating \cref{eq:interQ-PF} with respect to $\v Q$, we can
determine the root of
$\frac{\partial{\mathcal{G}(\v Q_{s-1},\v Q)}}{\partial\v Q}=0$ to obtain the
maximum of the intermediate function $\mathcal{G}(\v Q_{s-1},\v Q)$ at
iteration $s$. By doing this, we obtain (\cref{sec:derivQ})
\begin{align} \label{eq:Qmax}
 \v Q &=  f_{\v Q_{s-1}}( \v Q) = \frac{1}{K}\sum_{k=1}^K \left[\sum_{j=1}^{N_p}w_{k,\v
     Q_{s-1}}^{(j)}\left(\frac{1}{S_{j,i}}\sum_{i=1}^{Np}w_{k-1,\v Q}^{(i)}\ \gv\psi_{k}^{j,i}\v B_{k}^{j,i}\right)\right]
\end{align}  
where
$\gv \psi_k^{j,i}\triangleq\exp\{-\frac{1}{2}(\v x_{k}^{(j)}-\mathcal{M}(\v x_{k-1}^{(i)}))^T\v Q^{-1}(\v x_{k}^{(j)}-\mathcal{M}(\v x_{k-1}^{(i)})\},$
\vskip .2cm

$\v B_{k}^{j,i} \triangleq(\v x_{k}^{(j)}-\mathcal{M}(\v x_{k-1}^{(i)}))(\v x_{k}^{(j)}-\mathcal{M}(\v x_{k-1}^{(i)}))^T$ and $S_{j,i} \triangleq\sum_{i=1}^{Np}w_{k-1,\v Q}^{(i)}\gv \psi_k^{j,i}$.\vskip .2cm

We have to take into account that particles and weights indexed by $j$,
that is, $\v x_{k}^{(j)}$ and $ w_{k,\v Q_{s-1}}^j$ in \cref{eq:Qmax}, are computed in the
expectation step using $\v Q_{s-1}$ as the model error covariance matrix. On
the other hand, particles and weights indexed by $i$, that is,
$\v x_{k-1}^{(i)},w_{k-1, \v Q}^i$ should be computed with a filter that assumes $\v Q$
as the model error covariance matrix. However,  this $\v Q$ is the matrix to
be found in the maximization step.

Summarizing, the two steps of this EM algorithm using a particle filter at
iteration $s$ are

\begin{itemize}
\item {\bf E-Step}: Given $\v Q_{s-1}$, use a PF with a model error covariance $\v Q_{s-1}$ to calculate the weights and particles which are needed for the  intermediate function $\mathcal{G}(\v Q_{s-1},\v Q)$.

\item {\bf M-Step}: Solve equation \cref{eq:Qmax}. 
\end{itemize}  

The M-Step at iteration $s$ requires to solve an implicit equation for the
covariance matrix $\v Q$. In this work we propose to solve this equation using
a fixed point algorithm in $\v Q$. This fixed point algorithm require extra
iterations at each iteration of the EM algorithm, however the conducted
experiments showed that less than six iterations of the fixed point algorithm
are enough to satisfy the required stopping criteria.  The Banach fixed-point
theorem, or contractive mapping theorem, 
guarantees the existence (and uniqueness) of a fixed point  of certain
mappings (functions) defined on a complete metric space, as long as these mappings are contractive. It
is not possible to show analytically that the function involved in our
fixed point algorithm is contractive due to the nonlinearity in the dynamical model, and since we are not under the hypothesis
of the Banach fixed-point theorem we cannot guarantee that our algorithm converges to a fixed
point. What we can guarantee, based on empirical evidence, is that the
algorithm stops after a few iterations satisfying a stopping criteria based on
the Frobenius norm  defined
below. 

The pseudocode of the proposed algorithm is presented in \cref{alg:QFP}. Within this pseudocode,  PF($\v Q$) indicates that the
particles and corresponding weights are obtained
by using a particle filter with $\v Q$ as model error covariance
matrix. The algorithm evaluates the fixed point function, \cref{eq:Qmax}, to obtain the new estimate of the parameters.

The stopping
criteria for the Fixed Point Algorithm (FP) is defined
as either

$\rm{stop}_C(\v Q_{FP}, \v Q_{FP0})= \frac{ ||\v Q_{FP}-\v Q_{FP0}||_{F}}{||\v
  Q_{FP}||_{F}}$, where $||\cdot ||_{F}$
is the Froebenius norm, is smaller than a previously set threshold, or the maximum number of fixed point iterations is reached.

\begin{algorithm}[h]
  \caption{EM algorithm to estimate $\v Q$ using a PF and a fixed point algorithm}
  \label{alg:QFP}
    \begin{algorithmic}
\State{Given $\v Q_{EM}$, $\v x_{0}^{1:N_p}$, $w_{0}^{1:N_p}$, $\epsilon_{EM},\epsilon_{FP}$}
\State{iter$_{EM}=1$}      
\While{(iter$_{EM}$ $\le$ MaxiterEM)\textbf{and} (stop$_{EM}$ $> \epsilon_{EM}$)}
\State{\textbf{E-Step}:}
\State{Compute $\{\v x_k^{(j)}, w_{k,\v Q_{EM}}^{(j)}\}$ using PF($\v Q_{EM}$)}
\State{\textbf{M-Step}}:
\State{iter$_{FP} = 1$; stop$_{FP}=2 \epsilon_{FP}$}
\State{$\v Q_{FP0}= \v Q_{EM}$}
         \While{(iter$_{FP} \le$ Maxiter$_{FP}$) \textbf{and} (stop$_{FP}> \epsilon_{FP}$)}
\State{       $\v Q_{FP} = f_{\v Q_{EM}}(\v Q_{FP0})$ from \cref{eq:Qmax}}
\State{        Compute stop$_{FP}=\rm{stop}_{C}(\v Q_{FP},\v Q_{FP0})$}
\State {Compute $\{\v x_{k-1}^{(i)},w_{k-1,\v Q_{FP}}^{(i)}\}$ using PF($\v Q_{FP}$)}
\State{       iter$_{FP}=$iter$_{FP}+1$; $\v Q_{FP0}= \v Q_{FP}$}
           \EndWhile
\State{      Compute stop$_{EM}=$ stop$_C(\v Q_{EM},\v Q_{FP})$}
\State{         $\v Q_{EM}= \v Q_{FP}$} 
\State{         iter$_{EM}=$iter$_{EM}+1$}
\EndWhile
    \end{algorithmic}
\end{algorithm}

\section{Numerical experiments design}\label{sec:experiments}

In order to evaluate the capabilities and performance of the proposed
methodology, numerical experiments were designed using two different 
dynamical models $\mathcal{M}$, a univariate linear Gaussian model and the
Lorenz-96 model \cite{Lorenz96}.
For each of these models we conducted twin experiments with different
settings. We first generated a set of noisy
observations using the dynamical model with
known parameters. Then, using these synthetic observations and the same stochastic
dynamical model
we estimated the model error covariance $\v Q$ with the proposed algorithm and
compared the results with those obtained with some classical
methodologies. These  experiments are useful to assess the convergence and
performance of the proposed methodology.

\subsection{Linear model}
\label{sec:linear}

A one dimensional linear Gaussian state-space model is defined as
\begin{align}
        {{x}}_k = \nu{x}_{k-1} + \beta_k \label{eq:modelar1} \\
        {{y}}_k = {x}_{k} + \epsilon_k \nonumber
\end{align}      
where $\{ x_k\}_{k=0:K}, \{y_k\}_{k=1:K} \in \mathbb{R} $, $\beta_k\sim \mathcal{N}(0,\v Q)$, $\epsilon_k\sim \mathcal{N}(0,\v R)$, $\nu$ is the
autoregressive  coefficient and $\v Q,\v R $ are the error
variances. The implementation of the EM algorithm to estimate the parameters
of this  linear Gaussian model using the Kalman filter and
smoother was firstly discussed by Shumway and Stoffer in \cite{art:Shum_Sto_82},
whereas in \cite{shumwaysto} the same authors provide a more
detailed derivation of this implementation. Discussions of the convergence of
the EM algorithm for this model can be found in \cite{art:Douc_etal_14,art:Wei_etal_90, art:Wu_83}.
A  set of one-dimensional experiments was conducted by generating $K =100$ noisy observations
$\v y_{1:K}$  using the linear model \cref{eq:modelar1} with known parameter
values $\nu = 0.8$,$\,\v Q_{true}=\v I_1,\, \v R = \v I_1$. Then, using the same model, 
these synthetic observations, and an initial guess $\v Q_0$  sampled from a
uniform distribution $\mathcal{U}[0.5,1.5]$, the model error variance $\v Q$ was estimated by four different algorithms:

\begin{itemize}
\item EM + VMPF:  the
algorithm here proposed, that is a version of the EM algorithm for a particle
filter without the need of a  particle smoother. The particle filter used is
the variational mapping particle filter (VMPF) described in \cref{sec:MPF}. 
\item EM + SIR: the algorithm here proposed, coupled with a classical Sampling Importance Resampling (SIR) filter \cite{art:Arulampalam_etal_02}.
\item EM + KF + KS:  the classical EM algorithm based on the Gaussian
  assumption that requires a  Kalman filter and smoother as presented in \cite{shumwaysto}.
\item EM + EnKF+EnKS: a version of the EM algorithm in conjunction with the
  EnKFilter and ENKSmoother as presented in \cite{art:Pulido_etal_18} with $N_p = 50$ ensemble members.
\end{itemize}

The procedure was repeated independently 50 times for each algorithm in order
to have an empirical distribution of the estimators. That
means that once generated the ``true'' state, 50 independent sets of
observations $\{\v y_{1:K}\}$ were generated using the same model;$\,$for each set
of observations an independent initial
guess $\v Q_0$ was sampled from a uniform distribution
$\mathcal{U}[0.5,1.5]$. Then, using the same model each set of synthetic observations and corresponding initial guess, the
model error covariance $\v Q$ was estimated. With these results we can have an
approximation of the empirical distribution of the estimators.

As the algorithm here proposed is suitable to be used with any particle
filter, we tested the EM + PF algorithm with two different particle filters, namely the classical
Sampling Importance Resampling (SIR) filter with $N_p=1000$ particles (for a detailed explanation see
\cite{art:Arulampalam_etal_02}) and the recently proposed particle filter based on
optimal transportation and referred as Variational Mapping Particle Filter (VMPF)
 (\cref{sec:MPF} and \cite{art:Pulido_MPF_19}) with $N_p=20$ particles. 

 \begin{figure}[htbp]
  \centering
  \includegraphics[width=0.8\linewidth]{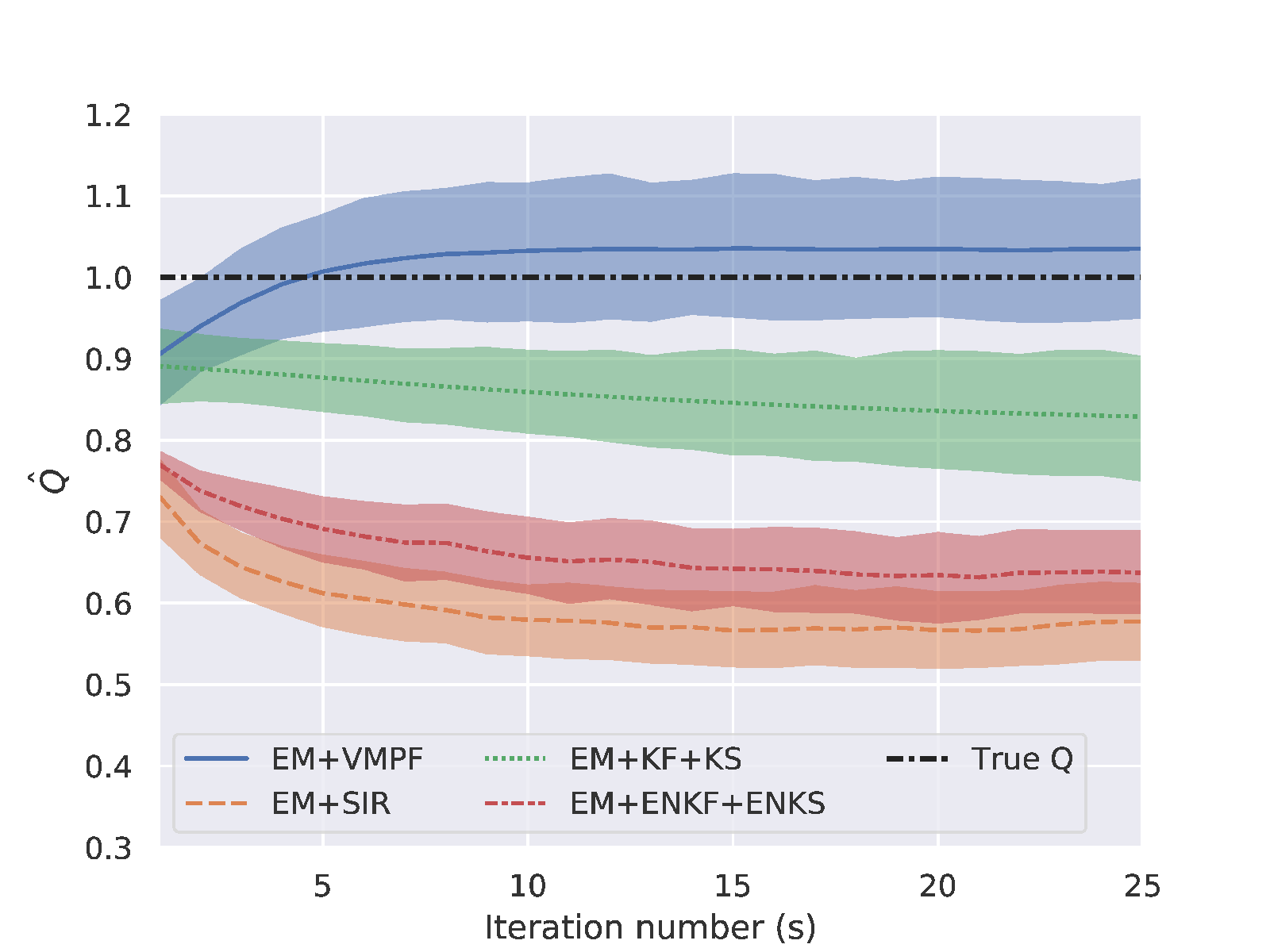}
  \caption{$\hat{\v Q}$ as function of the iteration number using the  EM + VMPF
    (blue), EM + SIR (orange) EM+KF+KS (green) and  EM + EnKF + EnKS (red)
    algorithms,  for the linear model \ref{eq:modelar1} with true parameter
    value $\v Q_{true} = \v I_1$, $\v R=\v I_1$, $\nu =0.8$ and $K=100$. The 95$\%$ confidence
    interval  were generated by running 50 independent repetitions of these
    estimation experiments and 25 EM iterations. For the EM + VMPF only $Np=20$ particles
  were used, whereas for the EM + SIR and the EM+EnKF+EnKS $N_p=1000$
  particles and $N_p=50$  ensemble members, respectively,  were used.}
  \label{fig:ar1R1}
\end{figure}

The mean and $95\%$ confidence intervals  obtained by the 50 repetitions of the experiments with each
algorithm  are shown in \Cref{fig:ar1R1}. 
 As can be seen in \Cref{fig:ar1R1}, although with a greater dispersion, the EM + VMPF algorithm starts to stabilize after as few as 4 iterations with its mean
value  very close to the true value $\v Q_{true}=\v I_1$. Despite having smaller variances and also
stabilizing after a few iterations, the other algorithms provide estimates that are more biased. Moreover, the
EM + EnKF + EnKS algorithm  reaches an estimated value of 0.65 which represents
a relatively large underestimation in coherence with the EM + KF + KS 
proposed in \cite{shumwaysto}.

 For this simple linear model and with $N_p$ = 1000 
 particles, the results obtained by using the SIR filter were similar to those
 obtained by the EM + EnKF + EnKS proposal, and not as good in
 terms of bias and RMSE as the ones obtained by using the VMPF with only
 $N_p=20$ particles. In the experiments that follow, we only use EM+VMPF for
 the particle filter implementation of the proposed algorithm. Experiments
 coupling the EM algorithm 
 with the SIR filter for high-dimensional state and observational spaces,
 including the Lorenz-96 system, are not feasible due to the large number of particles required \cite{art:Pulido_MPF_19}.

\begin{figure}[htbp]
  \centering
  \includegraphics[width=0.8\linewidth]{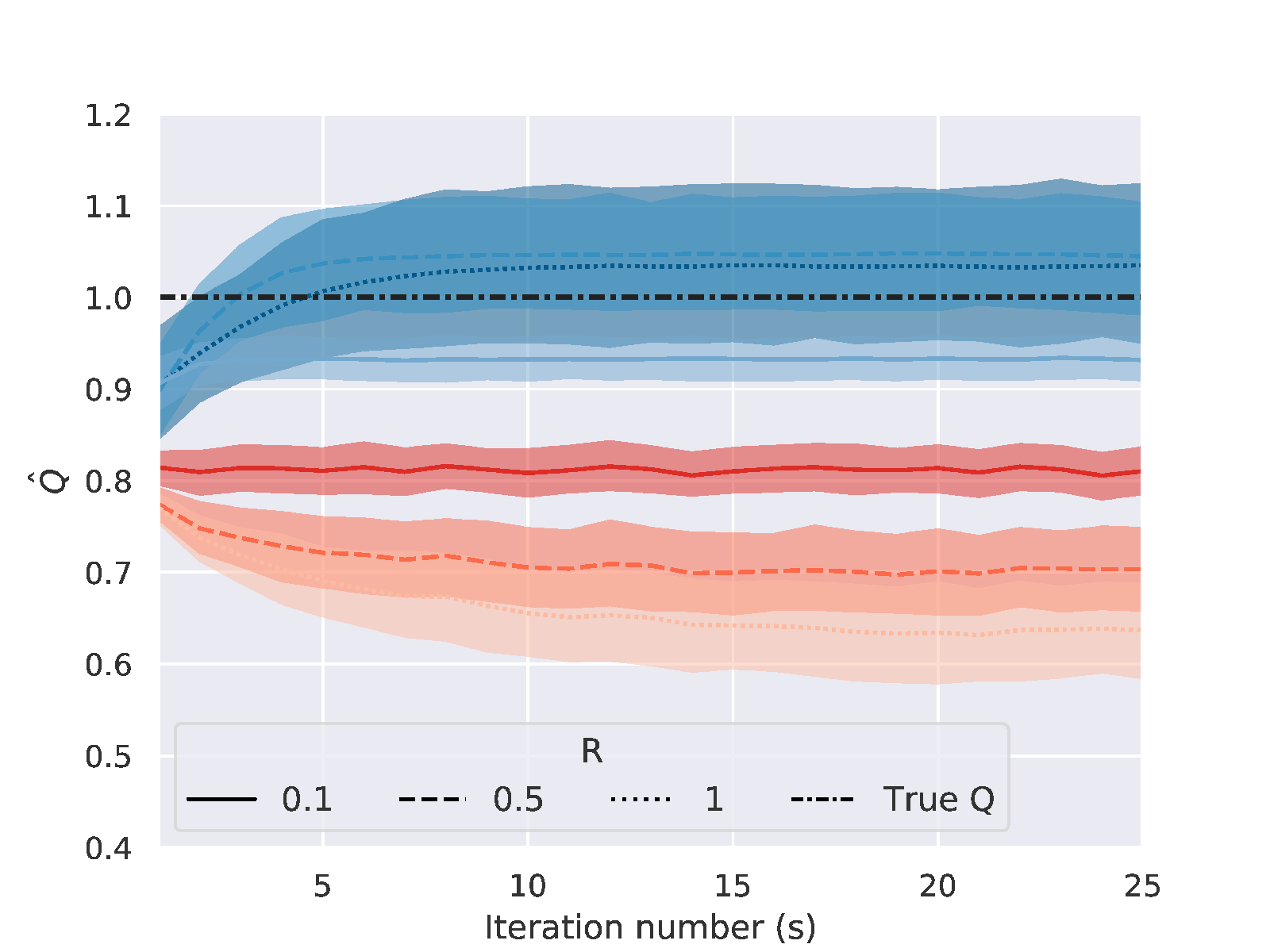}
  \caption{$\hat{\v Q}$ as a function of the iteration number using the EM+VMPF algorithm
    (blue), and the EM+KF+KS algorithm (red) for the linear model \ref{eq:modelar1}
    with true parameter values $\v Q_{true} = \v I_1$, $\nu =0.8$, different values of $\v R$
    and $K=100$.  For the algorithm here proposed (EM + VMPF) only $Np=20$
    particles were used, whereas for the EM+EnKF+EnKS $N_e=50$ ensemble members were used\ma{.}}
  \label{fig:ar1-cR}
\end{figure}

In order to  analize the sensitivity of the algorithm to different observation error
variances $\v R$, we conducted some experiments similar to the ones described above, for
different values of $\v R \in \{0.1\v I_1,0.5 \v I_1, \v I_1\}$. The estimates of the model error covariance $\v Q$ (mean value of 50 independent repetitions of
the same experiment) and $95\%$ confidence intervals as a function of the
iteration number of the EM + VMPF and EM + EnKF + EnKS algorithms are shown in
\cref{fig:ar1-cR}. Both algorithms stabilize in as few as 5 - 10 iterations,
with the EM + VMPF less biased, independent of the value
of $\v R$. Even for a small observation error variance $\v R$, the $95 \%$
confidence interval  obtained by the  EM + EnKF +
EnKS proposal does not reach the true value $\v Q = \v I_1$.

\subsection{Lorenz-96 model}

In this section we show results of twin experiments using the chaotic and
nonlinear Lorenz-96 system as the dynamical model $\mathcal{M}$ in
equation \ref{eq:model}. The Lorenz-96 is one of the most used toy models in data assimilation within the
geoscience community due to its ability to mimic certain properties of the
atmospheric predictability at a low computational cost.

It is defined by the ordinary differential equations

\begin{equation}
  \label{eq:lorenz96}
  \frac{\mathrm{d}X_n}{\mathrm{d}t} = -X_{n-2}X_{n-1}+X_{n-1}X_{n+1}-X_n+F,
\end{equation}  
where $X_n$ is the state variable of the model at variable (position) $n$, $n=1, \ldots, N_x$ and the domain is assumed periodic, that is, $X_{-1}\equiv X_{N-1}$,
$X_{0}\equiv X_{N}$, $X_{N+1}\equiv X_{1}$. $F$ is the forcing constant and as usual, here it is set $F=8$ to represent chaotic dynamics.

We used a fourth-order Runga-Kutta scheme, with a model time step of
$\delta_t= 0.005$ to integrate  the Lorenz-96 equations \cref{eq:lorenz96}. In
these first set of experiments with the Lorenz-96 system, the number of
variables is set to $N_x=8$, meaning that 64 parameters have to be  estimated as
the model error covariance $\v Q$ is an $8\times 8$ matrix. The observation error covariance $\v R$ was chosen as a diagonal matrix
$\v R = \sigma_R^2\mathrm{\bf I}$. We observed every grid point, in other
words, the observation model $\mathcal{H}$ is assumed to be the identity
transformation. The observations are taken every $\Delta_t = 0.05$ which represents 10 model time steps in all the experiments performed.

For these twin experiments, we simulated $K=500$ noisy observations $\v y_{1:K}$
using the Lorenz-96 model  with known model error and
observation error covariances $\v Q_{true}, \v R$. Using the same model and these
synthetic observations, the full model error covariance matrix $\v Q$ was
estimated.

The particle filter used in all the experiments performed was the VMPF with $N_p= 20$ particles. The results are compared with the ones from the EM + EnKF+EnKS algorithm with also $N_p=20$ ensemble members.

To assess the proposed methodology to
estimate the model error covariance matrix, experiments with two different structures, usually assumed in
practice, were proposed  for
the true model error covariance matrix $\v Q_{true}$: a) an isotropic non correlated
covariance matrix, where $\v Q_{true}= \sigma^2 \mathrm{\bf I}$, with $\mathrm{\bf I}$ 
the identity matrix of order $N_x$ and b) an isotropic tridiagonal covariance
matrix with 
diagonal values $\sigma_d^2$ and both  sub-/super-diagonal values
$\sigma_{sd}^2$. In the first case we assume
that model errors for different model variables are uncorrelated and have the
same variance $\sigma_d^2$, whereas for
the second case we assume an a priori spatial covariance structure with
correlations between the first neighbours.

Following a similar procedure as for the linear model case (\cref{sec:linear}),
$50$ independent realizations of this experiment were performed in order to show
the estimator empirical
distribution and to analyze its sensitivity to initial guesses $Q_{0}$ and random
sampling of the observations. The non zero values of the  initial guesses
$Q_{0}$ (a $N_x\times N_x$ matrix with the same structure of $\v Q_{true}$), $\sigma_{d}^2$ and $\sigma_{sd}$, were sampled from
$\mathcal{U}[0.05,0.5]$ and $\mathcal{U}[0.01,0.15]$, respectively.

\begin{figure}[htbp]
  \centering
  \includegraphics[width =0.8\linewidth]{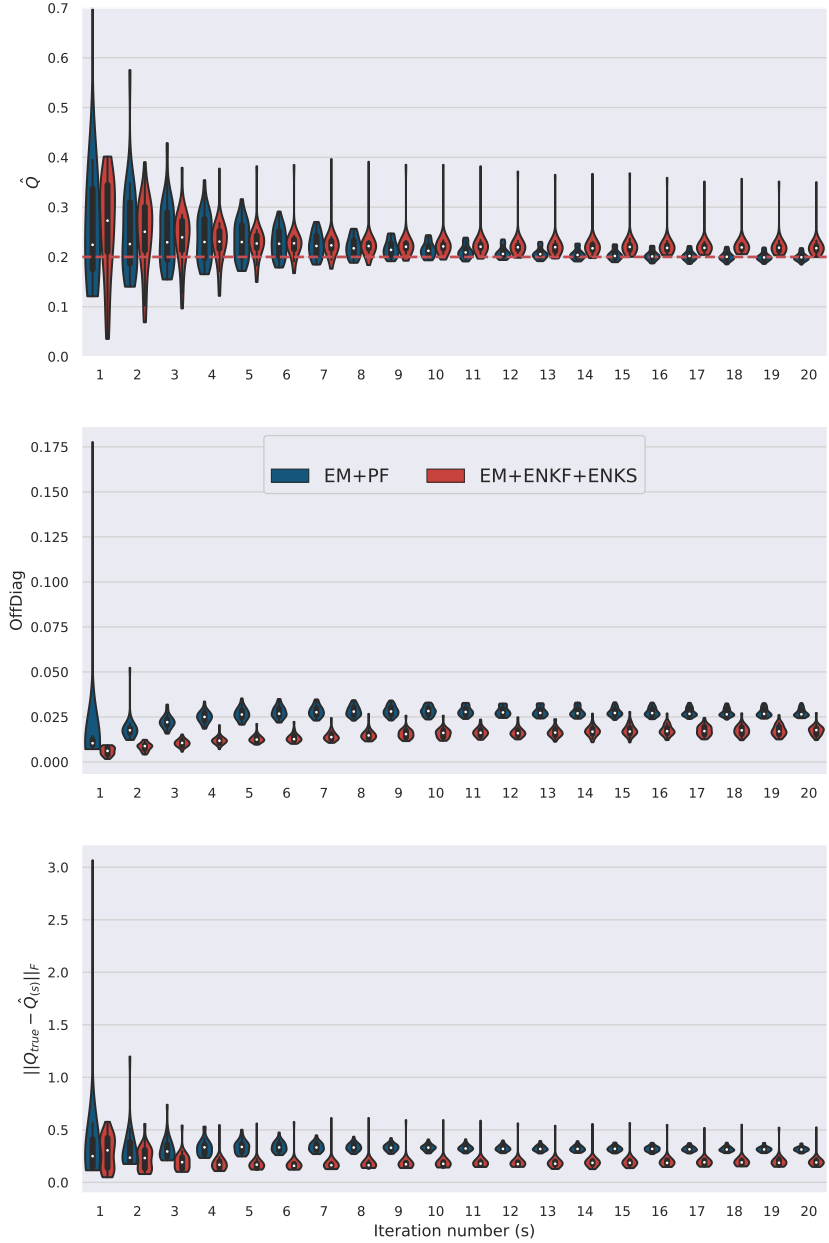}
  \caption{EM+ VMPF (blue) and EM + EnKF + EnKS (red) performances when estimating
    $\v Q$ (top), mean of the absolute value of the off-diagonal elements of
    $\hat{\v Q}_{(s)}$ and Frobenius norm  $\lVert\v Q_{true}- \hat{\v Q}_{(s)}\rVert_{F}$ and  loglikelihood function $l(\hat{\v Q}_{(s)})$ for the Lorenz-96 model \ref{eq:lorenz96} with
    true model error
    covariance  $\v Q_{true} = \sigma^2 \mathrm{I}$, with
    $\sigma^2=0.2$, and $K=500$. The violin plots were
    generated by running 50 independent repetitions of the algorithm  with
    $Np=20$ particles and keeping the average of the diagonal
  values of $\hat{\v Q}_{s}$}
  \label{fig:l96-8-Q0.2}
\end{figure}

In  \cref{fig:l96-8-Q0.2} we show the empirical distribution of the
estimator of $\v Q$ (top panel) for true parameter value
$\v Q_{true} =\sigma^2\mathrm{\bf I}$ with $\sigma^2=0.2$, $\v R=0.5\mathrm{\bf I}$ and $K=500$ observations, the distribution  of
the mean of the absolute values of the
off-diagonal elements of $\hat{\v Q}_{(s)}$ (middle panel) and the Frobenius
norm $\lVert\v Q_{true}- \hat{\v Q}_{(s)}\rVert_{F}$ (bottom panel), for the
algorithm proposed in this work, EM+VMPF (blue) and the  EM+EnKF + EnKS
algorithm (red)
proposed by \cite{art:Pulido_etal_18}, that requires an ensemble Kalman smoother.

For ease of visualization,  and just for plotting purposes,  in this case of an isotropic uncorrelated
covariance assumption $\v Q =\sigma^2 \mathrm{\bf I}$,  at the s-th iteration
of the EM algorithm  we kept $[\widehat{\sigma_d^2}]_{s}$ as the average of the diagonal values of
$\widehat{\v Q}_{(s)}$. As we repeated each experiment independently 50 times, we
have a series of 50 values of $[\widehat{\sigma_d^2}]_{s}$  at the $s$ iteration of the EM-algorithm to
construct a violin object that describes the empirical distribution of the
corresponding estimator at each iteration. 


With only $N_p =20$  particles both  methods provide good estimates of $\v Q$,
stabilizing in about 10 iterations  with its median
value (white circle) around the true diagonal value (top
panel). The EM+ VMPF (blue) algorithm produces estimates less biased than the
EM + EnKF + EnKS (red) despite the fact that it does not require a particle
smoother (and therefore uses less observational information in the state estimates). More over, the violin objects
show that the empirical distribution for the EM + VMPF estimates is symmetric
and highly concentrated around the true value, whereas 
the EM+EnKF+EnKS estimates empirical distributions have a greater dispersion,
are not symmetric and have tails towards higher values of $\sigma_d^2$,
meaning that in the 50 repetitions
performed this method  overestimates the value of $\sigma_d^2$.
The middle panel of \cref{fig:l96-8-Q0.2} shows the distribution  of
the mean of the absolute values of the
off-diagonal elements of $\hat{\v Q}_{(s)}$. Both methods provide reasonable
estimates of the off-diagonal elements (should be zero), with the
EM+EnKF+EnKS  better. The bottom panel of the same figure shows the
empirical distribution of the  Frobenius
norm $\lVert\v Q_{true}- \hat{\v Q}_{(s)}\rVert_{F}$ for both methods. Again,
the EM + EnKF + EnKS (red), despite having a greater dispersion has a slightly
better performance. 

\begin{figure}[htbp]
  \centering
  \includegraphics[width=0.9\linewidth]{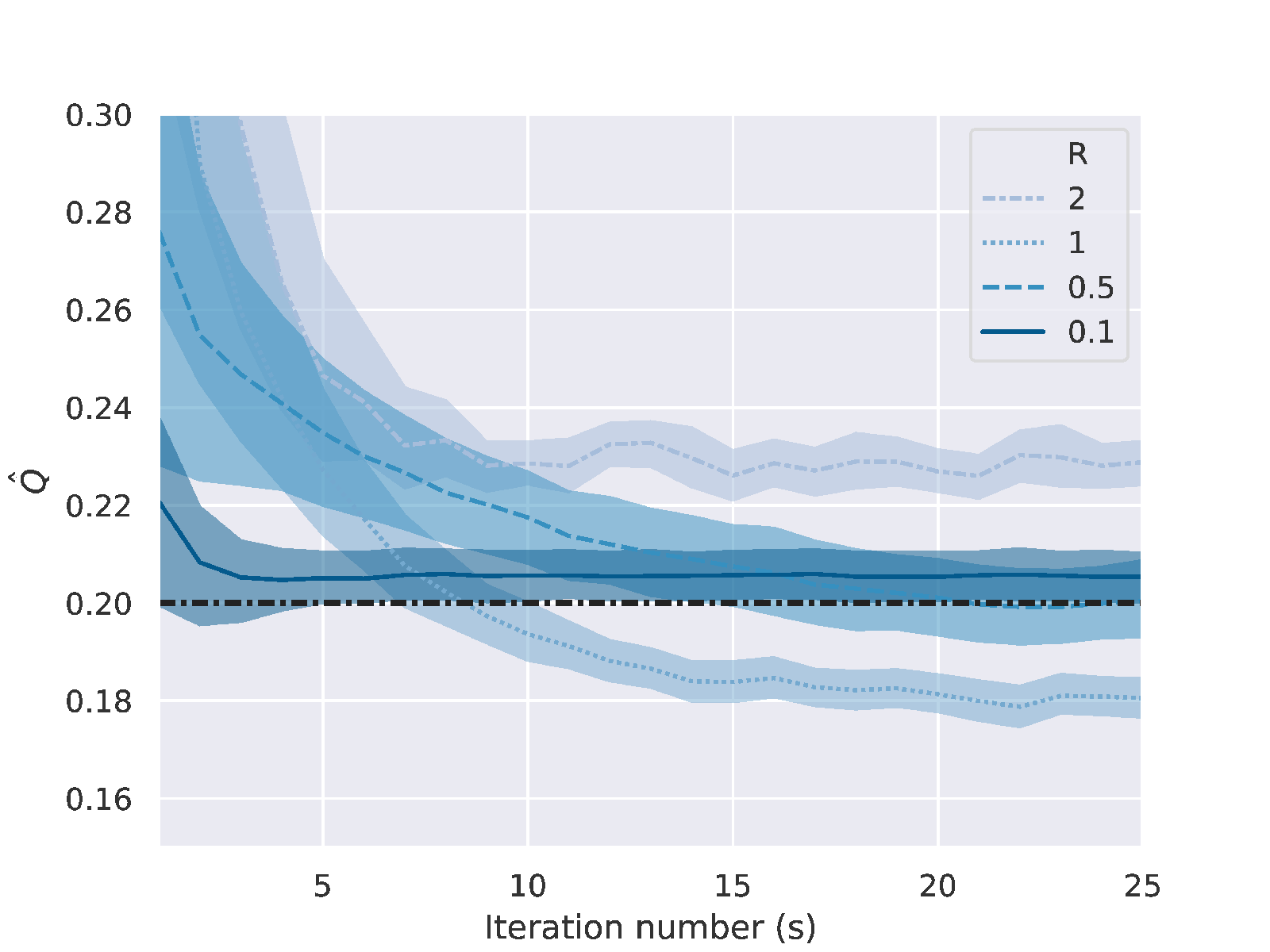}
  \caption{Sensitivity of EM+VMPF to different values of observation error $\v R$  with
    true model error
    covariance  $\v Q_{true} = \sigma^2 \mathrm{I}$, with
    $\sigma^2=0.2$, and $K=500$.}
  \label{fig:varyingR}
\end{figure}

To analyze the sensitivity of the
proposed algorithm to different values of the observation error covariance
$\v R$ a second set of experiments was performed. We assumed $\v R=\sigma_R^2\v I$ with $\sigma_R^2 \in \{0.1,0.5,1.0,2.0\}$, and $\v Q = 0.2\v I$. For each  setting we performed 30 independent realizations of
the same experiment. The mean and $95\%$ confidence intervals for 
$[\widehat{\sigma^2_{d}}]_{s}$ for each value of $\sigma_R^2$ are shown in
\cref{fig:varyingR}. The larger $\v R$ values, the harder to estimate model error likely because of sampling noise. The results for $\sigma_R^2=1.0$ and 2.0 are noisier and demonstrate 
small biases, even after 25 iterations of the EM scheme. As clearly stated in \cite{art:review_tandeo} and references therein, the quality of
reconstructed state vectors and estimation procedures when using variational
or ensemble-based methods, largely depends on the
relative amplitudes between observation and model errors. In
\cite{art:Zhu_etal_18} the authors also mention that a small increment in the
magnitude of the observation error $\v R$ highly affects the accuracy and
behaviour of the method  they propose to  estimate the model error $\v Q$
iteratively in the observation space using the implicit equally weighted
particle filter (IEWPF) \cite{art:Zhu_etal_16}. Their method provides reasonable estimates of the diagonal  values of $\v Q_{true}$ as long as the diagonal values of
the observation error matrix
$\v R$ are relatively small (they assume a diagonal $\v R$), with $\sigma_R^2=0.2$ 
the largest one. The larger $\sigma_R^2$, the less accurate the estimation
procedure. As shown in \cref{fig:varyingR}, the estimation with the EM+VMPF
gives results with a good performance for observational variances ranging from
0.1 to 2.0. The estimation error increases with the increase of the
observational variance, however, the estimation error is lower than 20\% in
all the cases shown.

\begin{figure}[htbp]
  \centering
  \includegraphics[width=0.8\linewidth]{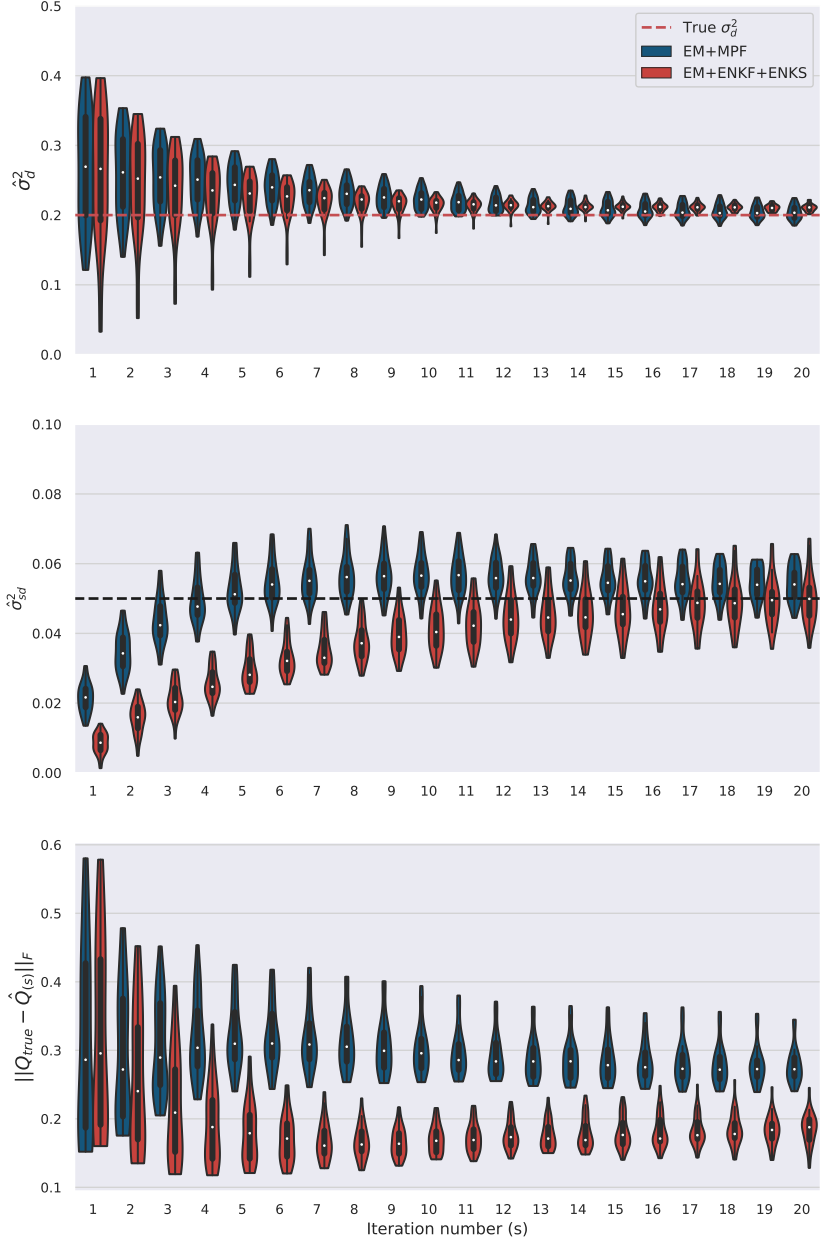}
  \caption{Estimation of  $\v Q$ as a
    function of the iteration for the EM+VMPF (blue) and EM+EnKF+EnKS (red)
    for the Lorenz-96 model with tridiagonal true model error covariance  $\v
    Q_{true}$, with   $\sigma_{d}^2=0.2$, $\sigma_{sd}^2=0.05$ and
    $K=500$. The violin plots were
    generated by running 50 independent repetitions of these algorithms  with
    $N_p=20$ particles. Upper panels show the average of the estimated diagonal
  values of $\hat{\v Q}_{s}$ (top panel) and the average of the
  sub/super-diagonal values of $\hat{\v Q}_{s}$ (middle panel). Frobenius norm $\lVert\v Q_{true} - \hat{Q}_{(s)}\rVert_{F}$
  (bottom panel) as function of the algorithm iteration.}
  \label{fig:l96-Qtri}
\end{figure}

Experiments designed for the estimation of a 
tridiagonal isotropic model error covariance are shown in
\cref{fig:l96-Qtri}. The diagonal value of $\v Q_{true}$ was set to
$\sigma_d^2 = 0.2$ and  sub-/super-diagonal values were $\sigma_{sd}^2 =0.05$,
as defined in \cite{art:Zhu_etal_18}, $\v R = 0.5 \mathrm{\bf I}$,  $K=500$,
$N_p=20$.
For this isotropic tridiagonal model error covariance assumption, at the s-th
iteration of the EM algorithm we computed $\widehat{\sigma^2_{d}}$ as the average
of the diagonal values of $\hat{\v Q}_{s}$, and $\widehat{\sigma^2_{sd}}$ as the average
of the sub-/super-diagonal values of $\hat{\v Q}_{s}$. The violin plots
at the s-th iteration of the EM algorithm were generated with 
$\widehat{\sigma^2_{d}}$, $\widehat{\sigma^2_{sd}}$ obtained 
for  each realization of the experiment.

As shown in \cref{fig:l96-Qtri} both the EM+VMPF(blue) and EM+EnKF+EnKS (red) methods converges rapidly to the
true value $\sigma_d^2$ (top panel). The empirical distribution of the
EM+VMPF estimates shows a median value (white circle) closer to the true value
of  $\sigma_d^2$ but has a greater dispersion. However, as in the diagonal
$\v Q$ case, the EM+EnKF+EnKS proposal tends to overestimate $\sigma_d^2$. Both
methods show a similar performance when estimating the  sub-/super-diagonal
elements of $\v Q$, with the EM+VMPF more biased in terms of the median
value. However, after 17 iterations the violin object  obtained by
the EM+VMPF proposal is completely contained within the violin object
obtained by the EM+EnKF+EnKS algorithm. A different behaviour is observed for
the empirical distribution of the Frobenius Norm. In this case, the
EM+EnKF+EnKS always outperforms the EM+VMPF estimation procedure.



In most of the experiments performed the proposed method converges to the
true diagonal value of $\v Q$ after a few iterations, tending to slightly
overestimate the sub-/super-diagonal values of $\v Q$. 

\begin{figure}[htbp]
  \centering
  \includegraphics[width=0.8\linewidth]{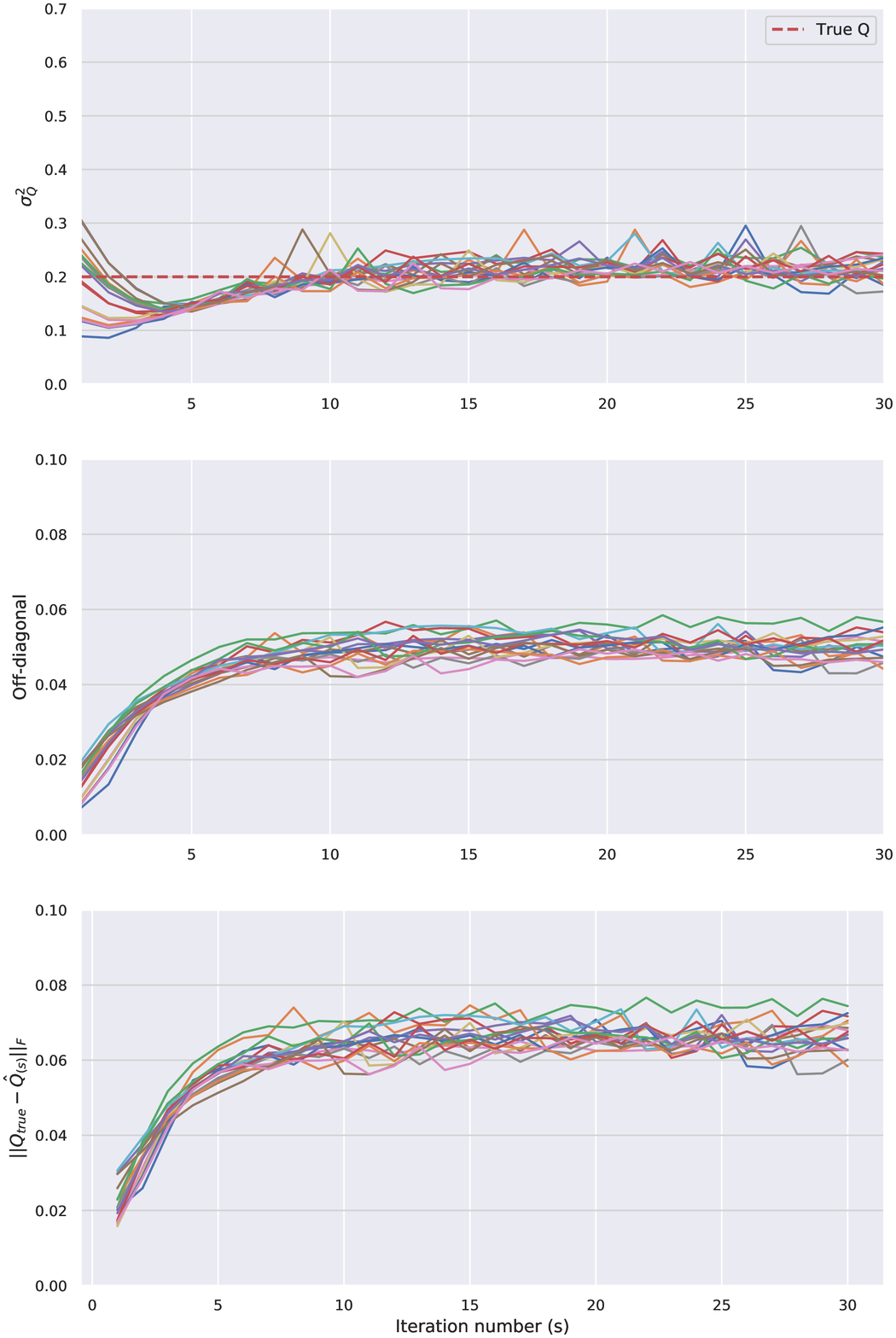}
  \caption{EM+VMPF results when estimating $\v Q$ as a
    function of the iteration, for the 40th dimensional Lorenz-96 model  with     $\v Q=0.2\ \v I$, $\v R=0.5\ \v I$, $K=250$. Twenty independent repetitions of  the same experiment were performed and each line shows the mean of the diagonal values of $\hat{\v Q}_{s}$ (top panel), mean of the absolute value  of the off-diagonal elements of $\hat{\v Q}_{(s)}$ (middle panel) and Frobenius norm $\lVert\v Q_{true}-\hat{\v Q}_{(s)}\rVert_{F}$ (bottom panel) for each of these repetitions.}
  \label{fig:L96-nx40}
\end{figure}

A higher dimensional experiment was also performed for a chaotic
Lorenz-96 model with 40 dimensions and $F = 8$. In this case, the number of parameters to
be estimated is $40\times 40 = 1600$. Estimations of model error covariances from 20 repetitions of this experiment
for an isotropic uncorrelated  model error covariance $\v Q = 0.2\ \v I$, with
$\v R =0.5\ \v I$ and K = 250 observations are shown in
\cref{fig:L96-nx40}. Despite the increase in the dimensionality, the
method provides unbiased results when estimating the diagonal values of $\v Q$ after a few
iterations (top panel), however the estimation of the off diagonal values and
Frobenius norm of $\v Q$ is slightly affected by the increase in
dimensionality if compared to the $8 \times 8$ case (cf. \cref{fig:l96-8-Q0.2}).

\section{Conclusions}\label{sec:conclusions}

In this work a novel method to estimate the model error
uncertainty in dynamical systems is introduced and evaluated. It  assumes that both the model and observation
errors are additive and Gaussian with zero mean and  covariance matrices $\v Q$ and $\v R$, respectively. The methodology here presented is based on the maximization
of a likelihood criterium using the principles of the EM algorithm and a
particle filter. We aim at
maximizing the complete likelihood of the observations by marginalizing this 
likelihood function. The resulting likelihood is expressed sequentially. By taking this approach, in the E-Step of the EM
algorithm we only 
have to compute filtering densities  avoiding the need to
compute smoothing densities, which are known
to be one of the main drawbacks when using particle filters in data
assimilation. The trade-off of avoiding the need of a particle smoother in the
E-Step of the EM algorithm is the need to solve an implicit equation for $\v
Q$ in  the M-Step of the EM algorithm. This problem was tackled by means of a
fixed point algorithm, and despite the fact that an analytical proof
of its convergence is not straightforward to obtain due to the nonlinearity of
the model dynamics, empirical results show that it converges to a
solution of this implicit equation.

The EM algorithm coupled with the VMPF presents, in general, an overall excellent performance.
It gives very promising results in
the experiments performed with a simple linear first order autoregressive
system and a chaotic Lorenz-96 system with 8 and 40 variables. In the first
case, results were compared with those obtained by different methods already
proposed in the literature, showing a good performance in terms of bias and
RMSE, and being more robust to different values of the observation error $\v R$. The new method is suitable for non-Gaussian posterior densities from nonlinear dynamical and observational models, unlike the Kalman Filter/Smoother
and its ensemble variants.

In the case of the Lorenz-96 system, its performance was tested for
different scenarios showing good convergence properties. It is stable even for
$\v R = 10\v Q$, although a small bias appears in the estimate.
The new method outperforms the traditional EM algorithm 
with the EnKS \cite{art:Dreano_etal_17} for a diagonal
$\v Q$ and for the diagonal of a tri-diagonal $\v Q$. However, off-diagonal
elements estimates were always noisier than those using an EnKS. 
It also  works in high-dimensional estimation problems of dimensions over 1000 and state space of 40.

Model error covariances are essential in particle flow filters. The conducted
experiments show that these particle flow filters, in particular VMPF, can
work with an adaptative model error without apriori information on this
covariance whereas in previous studies a fixed known model error covariance
was used \cite{art:Pulido_MPF_19}. Model error covariances impact on the prior
density and also on the kernel covariance in the VMPF. The overall excellent
performance of the estimates may also trace back to the strong sensitivity of
VMPF performance to model error covariance. In this sense, there is a positive feedback between the model error covariance estimates of the EM algorithm and the state estimates of the filter.

The computational cost of the algorithm here proposed is directly related to
the number of iterations needed for convergence. All the experiments performed
achieved convergence to a narrow neighbourhood of the true value of $\v Q$ in as few as 10 to 15
iterations of the EM algorithm. Each EM iteration requires the computation of
K filtering densities computed by using a particle filter with $N_p$
particles, whereas the M-Step requires to solve a fixed point algorithm. In
our experiments we set the number of iterations for this fixed point algorithm to
6, based on empirical evidence. In turn, each of these fixed point iterations also
require to compute K filtering densities computed by using a particle filter with $N_p$
particles.

\section{Appendix A: Derivation of $\mathcal{G}(\theta', \theta)$} \label{app}

 As explained in \cref{sec:inference}, the likelihood of the observations
can be decomposed as $p (\v y_{1:K};\theta) =\prod^K_{k=1}p(\v y_{k}|\v y_{k-1};\theta)$,
with the convention $\v y_{1:0}=\{\emptyset \} $. Marginalizing this last
expression we obtain
\begin{align} \nonumber
  p (\v y_{1:K};\theta) &=\prod^K_{k=1} p(\v y_{k}|\v y_{k-1};\theta)\\
  &= \prod^K_{k=1} \int   p(\v y_{k}|\v x_{k};\theta) p(\v x_{k}|\v y_{1:k-1};\theta) \mathrm{d}\v x_k
\end{align}

and taking logarithm we can rewrite this last expression as

\begin{align} \nonumber
l(\theta)=\log p (\v y_{1:K};\theta) = \log \prod^K_{k=1} \int q_{\theta'}(\v \v x_{k}) \frac{p(\v y_{k}|\v x_{k};\theta) p(\v x_{k}|\v y_{1:k-1};\theta)}{q_{\theta'}(\v \v x_{k})} \mathrm{d}\v x_k
\end{align}

where $q_{\theta'}(\v x_{k})$ is a probability density  function whose support includes the support of the likelihood of the
observations. In principle, $\theta'$ is
not necessarily equal to $\theta$ . Using Jensen's inequality,

\begin{align}
l(\theta) \ge \sum_{k=1}^K\int q_{\theta'}(\v x_{k}) \log\left(\frac{p(\v y_{k}|\v x_{k};\theta) p(\v x_{k}|\v y_{1:k-1};\theta)}{q_{\theta'}(\v x_{k})}\right) \mathrm{d}\v x_k \label{jensen-llik}
\end{align}

Let $\mathcal{G}(q,\theta)$ be an intermediate function defined as

\begin{align}
\mathcal{G}(q,\theta) = \sum_{k=1}^K\int q(\v x_{k}) \log\left(\frac{p(\v y_{k}|\v x_{k};\theta) p(\v x_{k}|\v y_{1:k-1};\theta)}{q(\v x_{k})}\right) \mathrm{d}\v x_k \label{itermFn}
\end{align}

Using Bayes rule, the recursive posterior density at time $k$ is
\begin{align}
p(\v x_k|\v y_{1:k};\theta)= \frac{p(\v y_{k}|\v x_{k};\theta) p(\v x_{k}|\v y_{1:k-1};\theta)}{p(\v y_k|\v y_{1:k-1};\theta)} \label{recPost}
\end{align}

If we choose $q(\v x_{k}) = p(\v x_{k}|\v y_{1:k};\theta)$, and so replacing \cref{recPost} in \cref{itermFn},it can
be shown that $\mathcal{G}(q,\theta)= l(\theta)$. That means that for
fixed $\theta=\theta'$,  the function $q$ that maximizes the
intermediate function, $\mathcal{G}$, is the recursive posterior density
$p(\v x_k|\v y_{1:k};\theta)$. This density can be inferred by a filter method
and this corresponds to the Expectation Step.

On the other hand, maximizing
$\mathcal{G}(q,\theta)$, with respect to $\theta$ gives a lower bound
of $l(\theta)$.

The maximization step consists in maximizing $\mathcal{G}(q',\theta)$ as a
function of $\theta$, where $q'$ is the density obtained in the
Expectation step.
If we now write $q_{\theta'}(\v x_{k}) = p(\v x_{k}|\v y_{1:k};\theta')$,
and make an abuse of notation in the expression
$\mathcal{G}(q_{\theta'},\theta)$ by replacing $q_{\theta'}$ by the parameter
that identifies it, then

\begin{align}
\mathcal{G}(\theta',\theta)&= \sum_{k=1}^K\int p(\v x_{k}|\v y_{1:k};\theta')\log\left(\frac{p(\v y_{k}|\v x_{k};\theta) p(\v x_{k}|\v y_{1:k-1};\theta)}{p(\v x_{k}|\v y_{1:k};\theta')}\right) \mathrm{d} \v x_k \label{itermFn2}\\
&= \sum_{k=1}^K\int p(\v x_{k}|\v y_{1:k};\theta')\log\left(p(\v y_{k}|\v x_{k};\theta) p(\v x_{k}|\v y_{1:k-1};\theta)\right) \mathrm{d} \v  x_k  \\
  &-\sum_{k=1}^K\int p(\v x_{k}|\v y_{1:k};\theta')\log\left( p(\v x_{k} |\v y_{1:k};\theta')\right) \mathrm{d} \v x_k 
\end{align}

\section{Appendix B: Equation for $\v Q$}\label{sec:derivQ}

We want to find the root of
$\frac{\partial}{\partial \v Q}\mathcal{G}(\v Q_{s-1}, \v Q)= 0$, where
$\mathcal{G}(\v Q_{s-1}, \v Q)$ 

\begin{align} \nonumber
         \mathcal{G}(\v Q_{s-1},\v Q) \doteq  \sum_{k=1}^K \sum_{j=1}^{N_p}w_{k, \v Q_{s-1}}^{(j)} \log \left(\sum_{i=1}^{N_p} w_{k-1, \v Q}^{(i)}\ \ \gv \phi\left(x_k^{(j)},\mathcal{M}(\v x_{k-1}^{(i)}),\v Q\right) \right)
\end{align}

and $\gv \phi\left(\v x_k, \gv \mu ,\gv \Sigma \right)=\frac{1}{(2\pi)^{N_x/2}|\gv\Sigma |^{1/2}}\exp\left\{-\frac{1}{2}(\v x_k -\gv \mu)^T\gv \Sigma ^{-1}(\v x_k -\gv \mu)\right\}$

\vskip.15cm

Denoting  $\gv\beta_k^{(i,j)}= \v x_k^{(j)} -\mathcal{M}(\v x_{k-1}^{(i)})$ we
have

\begin{align*}
  \frac{\partial}{\partial \v Q}\mathcal{G}(\v Q_{s-1},\v Q)& = \\
  = &\sum_{k=1}^K\sum_{j=1}^{N_p} w_{k,\v Q_{s-1}}^{(j)}\frac{\partial}{\partial \v Q}\log\left[\sum_{i=1}^{N_p}\frac{w_{k-1}^{(i)}}{(2\pi)^{N_x/2}|\v Q|^{1/2}}\times \right.\\
  & \left.\exp \left\{-\frac{1}{2}(\gv\beta_k^{(i,j)^T}\v Q^{-1}\gv  \beta_k^{(i,j)})\right\}\right]\\
 = &\sum_{k=1}^K\sum_{j=1}^{N_p} w_{k,\v Q_{s-1}}^{(j)}\frac{\partial}{\partial \v Q}\left[-\frac{N_x}{2}\log(2\pi)-\frac{1}{2}\log|\v Q|+ \right.\\
   &\left. +\log\left(\sum_{i=1}^{N_p}  w_{k-1}^{(i)}\exp\left\{-\frac{1}{2}(\gv \beta_k^{(i,j)^T}\v Q^{-1} \gv  \beta_k^{(i,j)})\right\} \right)\right]\\
&= -\frac{K}{2}\v Q^{-1}+\frac{1}{2}\v Q^{-1}\left[\sum_{k=1}^K\sum_{j=1}^{N_p}w_{k,\v Q_{s-1}}^{(j)}\ \frac{1}{S_i}\sum_{i=1}^{N_p}w_{k-1}^{(i)}\times\right.\\
   &\left.\times \exp\{-\frac{1}{2}(\gv \beta_k^{(i,j)^T}\v Q^{-1} \gv\beta_k^{(i,j)})\} \gv\beta_k^{(i,j)} \gv \beta_k^{(i,j)^T} \right]\v Q^{-1}
\end{align*}

\vskip.15cm
where $S_i= \sum_{i=1}^{N_p}w_{k-1}^{(i)}\exp\{-\frac{1}{2}(\gv \beta_k^{(i,j)^T}\v Q^{-1} \gv\beta_k^{(i,j)})\} $
\vskip.15cm

Thus,  $\v Q$ that satisfies  $\frac{\partial}{\partial \v Q}\mathcal{G}(\v Q_{s-1},\v Q)= 0$ is given by

\begin{align}
 \v Q &=  \frac{1}{K}\sum_{k=1}^K \left[\sum_{j=1}^{Np}w_{k,\v
     Q_{s-1}}^{(j)}\left(\frac{1}{S_{i}}\sum_{i=1}^{Np}w_{k-1,Q}^{(i)}\ \gv\psi_{k}^{j,i}\v B_{k}^{j,i}\right)\right]
\end{align}  
where

$\gv \psi_k^{j,i}=\exp\{-\frac{1}{2}(\gv \beta_k^{(i,j)^T}\v Q^{-1} \gv\beta_k^{(i,j)})\}$
  

$\v B_{k}^{j,i} =  \gv \beta_k^{(i,j)}\gv \beta_k^{(i,j)^T}$
\vskip .2cm

{\bf Acknowledgements:} This work was funded by European Research Council
(ERC) CUNDA project 694509 under the European Union Horizon 2020 research and innovation programme.

\end{document}